\documentclass[a4paper,11pt]{article}
\usepackage{pos}

\usepackage{dirtytalk}
\usepackage{mathtools,mathbbol}
\usepackage{graphicx,hyperref,xcolor}
\usepackage{float}
\def\Hc{\mathcal{H}}
\def\Pc{\mathcal{P}}
\def\Rc{\mathcal{R}}
\def\Tc{\mathcal{T}}
\def\Cc{\mathcal{C}}

\newcommand{\const}{\mathrm{const}}
\newcommand{\LF}{\left(}
\newcommand{\RF}{\right)}
\newcommand{\LT}{\left[}
\newcommand{\RT}{\right]}
\newcommand{\pd}{\partial}

\title{A robust explanation of CMB anomalies with
	a new formulation of inflationary quantum
	fluctuations}
\author*[a,b]{K. Sravan Kumar}
\author[c]{Jo\~ao Marto}

\affiliation[a]{Department of Physics, Tokyo Institute of Technology
	1-12-1 Ookayama, Meguro-ku, Tokyo 152-8551, Japan}

\affiliation[b]{Institute of Cosmology \& Gravitation,
	University of Portsmouth,
	Dennis Sciama Building, Burnaby Road,
	Portsmouth, PO1 3FX, United Kingdom}

\affiliation[c]{Departamento de F\'isica, Centro de Matem\'atica e Aplicações (CMA-UBI),
	Universidade da Beira Interior, 6200 Covilh\~a, Portugal}

\emailAdd{sravan.k.aa@m.titech.ac.jp,sravan.kumar@port.ac.uk}
\emailAdd{jmarto@ubi.pt}

\abstract{The presence of CMB Hemispherical Asymmetry (HPA) challenges the current understanding of inflationary cosmology which does not generically predict the parity violation in the primordial correlations. In this paper, we shall review the recently proposed resolution to this based on a new formulation of quantizing inflationary fluctuations by focusing on the discrete spacetime transformations in a gravitational context. The predictive power of   this formulation is that one can generate a scale dependent HPA in the context of single field inflation for all the primordial modes including scalar and tensor fluctuations without introducing any additional parameters. This result can be seen as an indication of spontaneous breaking of $\Cc\Pc\Tc$ symmetry in an expanding Universe, if confirmed by future observations it would be a great leap in the subject of quantum field theory in curved spacetime.}
\FullConference{%
  Corfu Summer Institute 2022 "School and Workshops on Elementary Particle Physics and Gravity",\\
  28 August - 1 October, 2022\\
  Corfu, Greece}

\begin{document}
\maketitle

\section{Introduction}
Inflationary cosmology by definition quasi-de Sitter (qdS) expansion phase in the early Universe and it has been the most successful paradigm in the early Universe cosmology
The predictions of scalar quantum fluctuations generated during inflation have been amazingly consistent with the latest Planck observations \cite{Starobinsky:1980te,Starobinsky:1979ty,Akrami:2018odb}. However, the latest observations from the Planck data along with its predecessor  NASA's Wilkinson Microwave Anisotropy Probe (WMAP) strongly indicate the presence of hemispherical power asymmetry (HPA) in the CMB sky \cite{Akrami:2019bkn} which is an anomaly that strongly questions the nature of inflationary quantum fluctuations \cite{Yeung:2022smn,Aluri:2022hzs}. The significance of HPA CMB is standing now at $3.3\sigma$ \cite{Akrami:2014eta}.  Being more precise, HPA means slightly more two-point temperature correlations  in the southern ecliptic hemisphere in comparison with the northern ecliptic hemisphere \cite{Creswell:2021eqi}.  
This anomaly has been found out to be significant at low-multipoles $\ell =2-64$ \cite{Schwarz:2015cma} ans less significant but still present up to the multipoles $\ell\sim 600$ according to the Planck 2013, Hansen \emph{et al.} and Aiola  \emph{et al.} \cite{Planck:2013lks,Hansen:2008ym,Aiola:2015rqa}. 
 The position dependent primordial power spectrum parametrization of the following, phenomenologically defines HPA as \cite{Lyth:2013vha,Lyth:2014mga}
\begin{equation}
	\Pc_\Rc\LF k,\,\hat{\boldsymbol{n}}\RF \simeq \Pc_{\Rc\:\: iso} (k)\LF 1+2A(k) \hat{\textbf{p}}\cdot\hat{\textbf{n}}\RF \,,
	\label{pwa}
\end{equation}
where $\Pc_{\Rc\:\: iso} (k)$ is the statistically homogeneous and isotropic power spectrum, $A(k)$ is the amplitude of the observed HPA which is constrained to be $\vert A \vert = 0.066\pm 0.021$ ($3.3\sigma$) at the large angular scales $\ell <64$ or at the wave numbers  $k \lesssim 0.0045\textrm{ Mpc}^{-1}$. Here  $\hat{\boldsymbol{p}}$ is the direction of maximal symmetry.  $\hat{\textbf{n}}=\frac{\boldsymbol{x}}{x_{\text{ls}}}$ is the line of sight from earth and $x_{\text{ls}} = 14,000 \textrm{Mpc}^{-1}$ is the comoving distance to the surface of last scattering. 
From \eqref{pwa} we can easily deduce \cite{Lyth:2013vha}
\begin{equation}
	A(k) = \frac{\Pc_\Rc\LF k,\,\hat{\textbf{n}} \RF - \Pc_\Rc\LF k,\,-\hat{\textbf{n}} \RF }{4\Pc_{\Rc\:\: iso}}\,. 
	\label{Akg}
\end{equation}

Quantum fluctuations during inflation are somewhat special because we need to quantize them in the qdS background. This means we would have to deal with quantum field theory in curved spacetime which is a subject that has not yet been fully understood. HPA strikingly show us the possible violation of parity which is a discrete transformation. As we know from the success of quantum field theory in Minkowski spacetime, the discrete symmetries play a very important role, right from the $\Cc\Pc\Tc$ (charge conjugation, parity and time reversal) invariance of scattering amplitudes to the notion of anti-particle which is a particle that propagates backward in time. In quantum theory time is a parameter whereas in classical General Relativity (GR) time is a coordinate and when the spacetime it self is dynamical and when one supposed to formulate a quantum theory in dynamical spacetime several conceptual conundrums are inevitable \cite{KSKJM}. The way forward would be a very careful observation of the meaning of discrete spacetime transformations in the gravitational context before defining quantum fields in a given spacetime. 

In this proceedings paper, we are going to review a new proposal for quantizing fields in an expanding Universe \cite{Kumar:2022zff} which explicitly defines the meaning of $\Cc\Pc\Tc$ invariance and its spontaneous symmetry breaking. In the later case one can precisely correlate it with the HPA of primordial power spectra. The crux of this investigation lies in distinguishing classical arrow of time (CAT) with the quantum mechanical arrow of time (QAT). The CAT is associated with for example the expanding Universe. CAT is also often called as thermodynamic arrow of time. The QAT on the other hand is intrinsic property of quantum theory where the initial and final conditions play a huge role. Just because Universe is found to be expanding, one cannot assume the quantum fields in the expanding Universe evolve in the same direction of time. In fact the expansion of Universe is only associated with growing scale factor rather than the notion of time itself. It is this subtle concept of time allow us to rethink on quantum fields in curved spacetime especially in the context of inflationary background. 

The paper is organized as follows. In Sec.~\ref{sec:dSDQFT} we discuss the de Sitter (dS) spacetime in the flat Friedmann-Lema\^itre-Robertson-Walker (FLRW) coordinates and the define the time reversal operation in the context of expanding Universe. We discuss the direct-sum quantum field theory (DQFT) formalism in dS and present analysis for the possible validity of $\Cc\Pc\Tc$ invariance in a language of two-point correlations. In Sec.~\ref{sec:DQFTinf} we present the definition of time reversal operation in qdS expansion with the expectation that $\Cc\Pc\Tc$ must be spontaneously breaking in qdS by the so-called slow-roll parameters. We then apply the DQFT to the inflationary flluctuations and compute the power spectra of scalar and tensor modes. 
In Sec.~\ref{sec:HPA} we 
explicitly witness the spontaneous breaking of $\Cc\Pc\Tc$ symmetry quantified by the amplitude of HPA which is of the order of slow-roll parameters especially at low-$\ell$. We will also see how for the first time we predict HPA for primordial gravitational wave spectra. In Sec.~\ref{sec:conc} we briefly conclude with highlighting the significance of results discussed in this paper and also comment on the future directions to explore. 

\section{Direct-sum  quantum field theory (DQFT)  in de Sitter spacetime}
\label{sec:dSDQFT}

The dS manifold is maximally symmetric spacetime like Minkowski and it is a solution of GR with a positive cosmological constant. Before we go to the context of quantum fields during inflation, it is important to understand the subject in dS. 
The dS spacetime is characterized the following relations between curvature tensors and the metric as
\begin{equation}
	R^{\mu}_{\nu\rho\sigma}  =\frac{R}{12}\LF \delta^\mu_\rho g_{\nu\sigma} -\delta^\mu_\sigma g_{\nu\rho}  \RF,\quad  R_{\nu\sigma} = \frac{R}{4}g_{\nu\sigma},\quad R=\const\,.
	\label{dSdef}
\end{equation}
The above definition is coordinate independent. In the context of cosmology the relevant form of expressing dS metric is in terms of 
 flat FLRW coordinates which is given by
\begin{equation}
	ds^2 = -dt^2 + a(t)^2d\textbf{x}^2
	\label{dsmetric}
\end{equation}
where the scale factor is 
\begin{equation}
	a(t) = e^{Ht},\quad H^2 = \LF\frac{1}{a}\frac{da}{dt}\RF^2>0\,, 
\end{equation}
Here $H$ is called the Hubble parameter. In this dS space, one has the comoving horizon defined by the radius
\begin{equation}
	r_H = \Big\vert  \frac{1}{aH} \Big\vert
	\label{horizonds}
\end{equation}
which decreases during expansion since scale factor grows. For the purpose of quantization it is useful to express \eqref{dsmetric} in terms of conformal time defined by
\begin{equation}
	d\tau = -\frac{dt}{a(t)}\,. 
\end{equation}
Integrating this equation, we obtain
\[
\tau= 	\frac{1}{a(t) H}
\,\text{,}
\]
Now the dS metric becomes conformal to Minkowski as
\begin{equation}
	ds^2 = \frac{1}{H^2\tau^2}\LF -d\tau^2+ d\textbf{x}^2 \RF
	\label{conmetric}
\end{equation}
When the time coordinate $\tau \gtrless 0$ we get a coordinate singularity at $\tau =0$ but it is just an artifact of the coordinate transformation. In reality, There is no spacetime singularity because all the curvature invariants in dS are constants. Just like Minkowski spacetime the dS metric satisfies $\Pc\Tc$ symmetries as
\begin{equation}
	\Pc\Tc: \tau\to - \tau,\quad \textbf{x} \to -\textbf{x}\,. 
	\label{disflat}
\end{equation}
We can notice that the time ($\tau$) reversal operation actually leads to time reversal in cosmic time, as well as the change of sign for the Hubble parameter $H$ i.e., 
\begin{equation}
	\Tc:	\tau \to -\tau \implies t\to -t,\, H\to -H\,. 
\end{equation}
We can verify that $H\to -H$ operation does not effect at all the definition of dS \eqref{dSdef}. For example, the Ricci scalar $R=12H^2$ in dS is completely symmetric under $H\to -H$. Remember that time is a parameter in quantum theory and here $\tau$ plays the role as a parameter for quantum field theory in dS. Since the metric \eqref{conmetric} is completely time symmteric (i.e., $\tau\to -\tau$) quantum fields cannot see the difference between whether $\tau$ is positive or negative. 
Indeed from \eqref{dsmetric}, we can notice that
\begin{equation}
	{\rm Expanding\,Universe:} \implies \begin{cases}
		t: -\infty \to +\infty,\quad  H>0 & (\tau: +\infty \to 0)\\ 
		t: +\infty \to -\infty,\quad  H<0 & (\tau: -\infty \to 0)
	\end{cases}
	\label{expcon}
\end{equation}
In \eqref{expcon} we see two arrows of cosmic time $t$ (together with sign of $H$), both of which correspond to expanding Universe with the definition that the $r_H$ decreases as scale factor increases \eqref{horizonds}. Therefore, \eqref{expcon} exactly serves the definition of time reversal in an expanding Universe and this is the cue for our formulation of direct-sum quantum field theory (DQFT). 

Let us start with direct-sum quantization of a massless scalar field whose action for the metric \eqref{dsmetric} can be computed as
\begin{equation}
	S_{\phi} = -\frac{1}{2}\int d\tau d^3x a^2	\phi\LF \pd_\tau^2+2\Hc \pd_\tau + k^2 \RF \phi\,, 
\end{equation}
where $\Hc = \frac{1}{a}\frac{da}{d\tau}=aH$. 
The above action is invariant under the $\Pc\Tc$ defined in \eqref{disflat}. 
To quantize it is convenient to rescale the field $\phi\to a \phi$,  which bring us to a harmonic oscillator form with time dependent mass
\begin{equation}
	S_\phi = \frac{1}{2} \int d\tau d^3x \Big[ \phi^{\prime 2} - \LF \pd_i\phi \RF^2  -  \frac{2}{\tau^2}  \phi^2 \Big]\,. 
	\label{msaction}
\end{equation}
In the process of quantization, we promote the scalar field $\phi$ as an operator $\hat{\phi}\LF \tau,\, \textbf{x} \RF$ which in DQFT formulation expressed as direct-sum \cite{Conway}
\begin{equation}
	\hat{\phi}\LF \tau,\, \textbf{x} \RF =  \frac{1}{\sqrt{2}}	\hat{\phi}_I\LF \tau,\, \textbf{x} \RF \oplus 	\frac{1}{\sqrt{2}}\hat{\phi}_{II}\LF -\tau,\, -\textbf{x} \RF = \frac{1}{\sqrt{2}}\begin{pmatrix}
		\hat{\phi}_I\LF \tau,\,\textbf{x} \RF & 0 \\ 0 & \hat{\phi}_{II}\LF -\tau,\,-\textbf{x} \RF
	\end{pmatrix} \,. 
	\label{disumdS}
\end{equation}
where 
\begin{equation}
	\begin{aligned}
		\hat{\phi}_I\LF \tau,\, \textbf{x} \RF &   = \frac{1}{\LF 2\pi \RF^{3/2}}\int  d^3k\Bigg[ \hat{c}_{I\,\textbf{k}} {\phi}_{I\,k}\LF \tau \RF e^{-i\textbf{k}\cdot \textbf{x}} + \hat{c}_{I\,\textbf{k}}^\dagger {\phi}^\ast_{I\,k}\LF \tau \RF e^{i\textbf{k}\cdot \textbf{x}} \Bigg]\, \\ 
		\hat{\phi}_{II}\LF -\tau,\, -\textbf{x} \RF &   = \frac{1}{\LF 2\pi \RF^{3/2}}\int  d^3k\Bigg[ \hat{c}_{II\,\textbf{k}} {\phi}_{-\,k}\LF -\tau \RF e^{i\textbf{k}\cdot \textbf{x}} + \hat{c}_{II\,\textbf{k}}^\dagger {\phi}^\ast_{II\,k}\LF -\tau \RF e^{-i\textbf{k}\cdot \textbf{x}} \Bigg]\,, 
	\end{aligned}
	\label{vdfieldS}
\end{equation}
The creation and annihilation operators here $\hat{c}_{\pm\textbf{k}},\, \hat{c}^\dagger_{\pm\textbf{k}}$ satisfy the following commutation relations 
\begin{equation}
	\begin{aligned}
		[\hat{c}_{I\,\textbf{k}},\,\hat{c}_{I\,\textbf{k}^\prime}^\dagger] & = 	[\hat{c}_{II\,\textbf{k}},\,\hat{c}_{II\,\textbf{k}^\prime}^\dagger] = \delta^{(3)}\LF \textbf{k}-\textbf{k}^\prime \RF\\
		[\hat{c}_{I\,\textbf{k}},\,\hat{c}_{II\,\textbf{k}^\prime}] &=	[\hat{c}_{I\,\textbf{k}},\,\hat{c}_{II\,\textbf{k}^\prime}^\dagger] = [\hat{c}_{I\,\textbf{k}}^\dagger,\,\hat{c}_{II\,\textbf{k}^\prime}^\dagger]  =0\,.
	\end{aligned}
	\label{comcandS}
\end{equation}
which leads to
\begin{equation}
	\begin{aligned}
		[\hat{	\phi}_I\LF t, \textbf{x} \RF,\,\pi_I\LF t,\,\textbf{x}^\prime \RF] &  = i \delta\LF \textbf{x}-\textbf{x}^\prime \RF\\
		[\hat{	\phi}_{II}\LF- t, -\textbf{x} \RF,\,\pi_{II}\LF -t,\,-\textbf{x}^\prime \RF] & = -i \delta\LF \textbf{x}-\textbf{x}^\prime \RF \\
		\LT \hat{\phi}_I\LF \tau,\,\textbf{x} \RF,\,\hat{\phi}_{II}\LF -\tau,\,-\textbf{x} \RF  \RT & = 0\,,
		\label{canDQFTds}
	\end{aligned}
\end{equation}
where 
\begin{equation}
	\pi_I \LF \tau,\,\textbf{x} \RF = \pd_\tau \phi_I,\quad  	\pi_{II} \LF -\tau,\,-\textbf{x} \RF = -\pd_\tau\phi_{II}\,.  
\end{equation}
The physical meaning of DQFT is the following. In the dS with flat FLRW we demand the a quantum field within the comoving horizon evolves forward in time at position $\textbf{x}$ while it evolves backward in time at the position $-\textbf{x}$.  The third line of \eqref{canDQFTds} is the most crucial one which ensures locality and causality of our (D)QFT framework. We can notice here that the field operator $\hat{\phi}_{II}$ canonical commutation relation in the second line of  \eqref{canDQFTds} differs from the first line by the sign of $i \to -i$ which indicates the time reversal $\tau \to -\tau$ because time reversal is an anti-Unitary operation in quantum theory. 
 In fact, in DQFT we divide the spatial region into two parts by parity, lets call them as region I and II. A quantum field in region one evolves forward in time and in the region II evolves backward in time in two different Fock spaces called ${\rm dS}_I$ and ${\rm dS}_{II}$ whose direct-sum forms the total Fock space of dS. 

The mode functions $\phi_{I\,k},\,\phi_{II\,k}$ are
\begin{equation}
	\begin{aligned}
		\phi_{I,\,k}  & = \alpha_{I\,k} \frac{e^{ik\tau}}{\sqrt{2k}}\LF 1+\frac{i}{k\tau} \RF +\beta_{I\,k} \frac{e^{-ik\tau}}{\sqrt{2k}}\LF 1-\frac{i}{k\tau} \RF\,, \\
		\phi_{II,\,k}  & = \alpha_{II\,k} \frac{e^{-ik\tau}}{\sqrt{2k}}\LF 1-\frac{i}{k\tau} \RF +\beta_{II\,k} \frac{e^{ik\tau}}{\sqrt{2k}}\LF 1+\frac{i}{k\tau} \RF\,, 
		\label{f1}
	\end{aligned}
\end{equation}
which satisfy $\phi_{I,\,k} \Big\vert_{\Tc:\,\tau \to -\tau} = \phi_{II,\,k}$ and are obtained by solving the Mukhanov-Sasaki equation \cite{Kumar:2022zff}
\begin{equation}
	\begin{aligned}
		\phi_{m\,k}^{\prime\prime}+ \LF k^2-\frac{2}{\tau^2} \RF \phi_{m\,k}=0\,. 
	\end{aligned}
	\label{MSeq}
\end{equation}
where $m= I,\,II$. In the limit $k^2\gg 2/\tau^2$ or $k^2\gg 2/r_H^2$, the mode functions $\phi_{m\,k}$ approach to the ones of Minkowski spacetime which is expected because at short distance scales the effects of curvature of spacetime should be negligible (See \cite{KSKJM} for more details). 
This fixes the Bogoliubov coefficients as 
\begin{equation}
\LF \alpha_{I\,k},\, \beta_{I\,k}\RF = \LF \alpha_{II\,k},\, \beta_{II\,k}\RF  = \LF 1,\,0 \RF\,, 
\end{equation}
which are compatible with the Wronskian conditions obtained from \eqref{canDQFTds}
\begin{equation}
	\begin{aligned}
\phi_{I,\,k} 	\phi^{\prime\ast}_{I,\,k}-	\phi^\ast_{I,\,k}	\varphi^\prime_{I,\,k} & = i \\	\phi_{II,\,k} 	\phi^{\prime\ast}_{II,\,k}-	\phi^\ast_{II,\,k}	\phi^\prime_{II,\,k} & = -i\,.
\end{aligned}
\end{equation}
The dS spacetime vacuum is now represented by the direct-sum
\begin{equation}
	\vert 0\rangle_{dS} = \frac{1}{\sqrt{2}}\LF\vert 0\rangle_{\rm dS_{I}} \oplus \vert 0 \rangle_{\rm dS_{II}}\RF = \frac{1}{\sqrt{2}} \begin{pmatrix}
		\vert 0 \rangle_{\rm dS_{I}} \\ 
		\vert 0 \rangle_{\rm dS_{II}}
	\end{pmatrix}\,.
	\label{dsumFds}
\end{equation}
where $	\vert 0 \rangle_{\rm dS_{I}},\,\vert 0 \rangle_{\rm dS_{II}}$ are 
vacuums defined by
\begin{equation}
	c_{I\,\textbf{k}}\vert 0 \rangle_{\rm dS_{I}} = 0,\quad c_{II\,\textbf{k}}\vert 0\vert 0 \rangle_{\rm dS_{II}} = 0
\end{equation}
We can verify this, observing that 
\begin{equation}
	\begin{aligned}
		& \frac{1}{a^2}{}_{dSI}\langle 0\vert \hat{\phi}_{I}\LF \tau,\, \textbf{x} \RF \hat{\phi}_{I}\LF \tau,\, \textbf{x}^\prime \RF\vert 0\rangle_{dSI} = \\ & \frac{1}{a^2} {}_{dSII}\langle 0\vert \hat{\phi}_{II}\LF -\tau,\, -\textbf{x} \RF \hat{\phi}_{II}\LF -\tau,\, -\textbf{x}^\prime \RF\vert 0\rangle_{dSII} = \frac{H^2}{4\pi^2k^3}\,. 
	\end{aligned}
	\label{eqcorrdS}
\end{equation}
This \eqref{eqcorrdS} implies that the correlations of quantum fields related by $\Pc\Tc$ transformations are identical in the case of dS spacetime.   Being more precise, quantum field correlations exiting the comoving horizon at the angular coordinates $\LF \theta,\,\varphi \RF$ (of the sphere of radius $r_H$ defined in \eqref{horizonds} ) are exactly the same as those exiting at the angular coordinates $\LF \pi-\theta,\,\pi+\varphi \RF$.  
This indicates in dS spacetime DQFT  formulation $\Pc\Tc$ or $\Cc\Pc\Tc$ (if we include charged fields) symmetry holds. In the case of qdS or inflationary spacetime we will see that the equality of correlations in the spatial regions divided by parity does not hold.  We shall discuss this in the next section. We note here that in a soon upcoming work more details of achieving Unitarity and observer complimentarity in dS DQFT can be found \cite{KSKJM}.

\section{DQFT in inflationary spacetime}
\label{sec:DQFTinf}

In the previous section, we have discussed quantization in dS spacetime by clear understanding of time reversal operation. In the context of inflationary spacetime time reversal would be much more involved than dS. As we learned in the previous section, the conformal time $\tau$ plays the role of parameter in quantum theory and $\tau\to -\tau$ actually mean time reversal in an expanding Universe for which we would have to change $t\to -t$ and $H\to -H$. This implies, if we want to understand time reversal in the context of curved spacetime or when there is gravity we would need additional parameters associated with gravity or background curved spacetime. For example, the Hubble parameter in the context of dS is the additional parameter along with cosmic time (of course together we can just use $\tau$ as a parameter). In the context of dynamical gravity like inflation, we would expect more parameters to play the role in time reversal. This is a very subtle task but however, with the generic expectation of $\Cc\Pc\Tc$ invariance must be spontaneously broken in curved spacetime we define the time reversal transformations in the inflationary context as  
\begin{equation}
	t\to -t \implies H\to -H,\quad \epsilon\to -\epsilon,\quad \eta\to -\eta\,. 
	\label{timerevqds}
\end{equation}
where $\epsilon = -\dot{H}/H^2, \eta = \dot{\epsilon}/(H\epsilon)$ are although usual slow-roll parameters, here in the context of time reversal their role is completely quantum mechanical and there is no classical analog (See \cite{Kumar:2022zff} for more details, also see  \cite{Rovelli:2004tv} for more thought provoking discussions on time reversal in quantum gravity). 

Similar to the dS DQFT, we should split the field as direct-sum of the two components as
\begin{equation}
	\hat{v}\LF \tau,\, \textbf{x} \RF = \frac{1}{\sqrt{2}} \hat{v}_{\rm +}\LF \tau,\, \textbf{x} \RF \oplus  \frac{1}{\sqrt{2}} \hat{v}_{\rm -}\LF -\tau,\, -\textbf{x} \RF = \frac{1}{\sqrt{2}}\begin{pmatrix}
		\hat{v}_{+}\LF \tau,\,\textbf{x} \RF & 0  \\ 0 & \hat{v}_{-}\LF -\tau,\,-\textbf{x} \RF
	\end{pmatrix}\,. 
\end{equation}
where $\hat{v}$ is the canonical Mukhanov-Sasaki variable corresponding to the only propagating scalar in the context of single field inflation. Here $\hat{v}_+\LF \tau,\,\textbf{x} \RF$ is the component when it acts on the corresponding vacuum it gives the component of the quantum field which evolves forward in time and similarly $\hat{v}_{-}\LF -\tau,\,-\textbf{x} \RF$ when it acts on its corresponding vacuum gives the component of the quantum field which evolves backward in time. 
The total vacuum of inflationary spacetime is the direct-sum of the two represented as
\begin{equation}
	\vert 0\rangle_{\rm qdS} = \vert 0 \rangle_{\rm qdS_{I}} \oplus \vert 0\rangle_{\rm qdS_{II}} = \frac{1}{\sqrt{2}}\begin{pmatrix}
		\vert 0 \rangle_{\rm qdS_{I}} \\ \vert 0\rangle_{\rm qdS_{II}}
	\end{pmatrix}\,. 
\end{equation}
The field operators $\hat{v}_{\pm}$ can be expanded as
\begin{equation}
	\begin{aligned}
		& \hat{v}_{\pm }=   
		\int \frac{d\tau d^3k}{\LF 2\pi \RF^{3/2}} \Bigg[ c_{\LF\pm \RF\textbf{k}} {v}_{\pm,\,k} e^{\mp i\textbf{k}\cdot \textbf{x}} + c_{\LF\pm \RF\textbf{k}}^\dagger {v}_{\pm,\,k}^\ast e^{\pm i\textbf{k}\cdot \textbf{x}} \Bigg]
	\end{aligned}
	\label{vid}
\end{equation}
with $c_{\LF\pm \RF\textbf{k}},\, c_{\LF\pm \RF\textbf{k}}^\dagger$ being the creation and annihilation operators the define the vacua as 
\begin{equation}
	c_{\LF + \RF\textbf{k}}\vert 0\rangle_{\rm qdS_I} = 0\quad c_{\LF - \RF\textbf{k}}\vert 0\rangle_{\rm qdS_{II}} = 0
	\label{qds1}
\end{equation}
The mode functions $ {v}_{\pm,\,k}$ are obtained by solving Mukhanov-Sasaki (MS) equations
\begin{equation}
	v_{\pm,\,k}^{\prime\prime}+ \LF k^2-\frac{{\nu}_s^{\LF \pm\RF 2}-\frac{1}{4}}{\tau^2} \RF v_{\pm,\,k}^2 =0\,.
	\label{MS-equation}
\end{equation}
where
\begin{equation}
	\nu_s^{\pm} \approx \frac{3}{2}\pm\epsilon\pm\frac{\eta}{2}
\end{equation}
We require that 
\begin{equation}
	\begin{aligned}
\big[c_{\LF + \RF\textbf{k}},\, c_{\LF - \RF\textbf{k}^\prime}\big]  = 0,\quad \big[c^\dagger_{\LF + \RF\textbf{k}},\, c^\dagger_{\LF - \RF\textbf{k}^\prime}\big]  = 0,\quad  \big[c_{\LF + \RF\textbf{k}},\, c^\dagger_{\LF - \RF\textbf{k}^\prime}\big]=0\,.  
	\end{aligned}
\end{equation}
This implies 
\begin{equation}
	\Big[\hat{v}_{+}\LF \tau,\, \textbf{x} \RF,\,\hat{v}_{-}\LF -\tau,\,-\textbf{x}^\prime \RF\Big] = 0
\end{equation}

%The evolution of the mode function $v_{-,\,k}$ is determined by solving the corresponding time reversed MS-equation \eqref{MS-equation}, for $\tau: +\infty \to 0$, and with a different $\nu_s^-$, such that

%Assuming $0<\LF \epsilon,\, \eta\RF \ll 1$ and constant we obtain  
\begin{equation}
	\begin{aligned}
		&	{v}_{\pm ,\,k}   = \frac{\sqrt{\pm \pi \tau}}{2} e^{\LF i\nu_s^{\pm}+1\RF} \Bigg[C_k^{\pm} H^{(1)}_{\nu_s^{\pm}}\LF \pm k \tau \RF
		+ D_k^{\pm} H^{(2)}_{\nu_s^{\pm}}\LF \pm k  \tau \RF\Bigg].
		\label{new-vac1}
	\end{aligned}
\end{equation}
The above mode functions can define the creation of positive frequency modes in the limit $\tau\to \pm\infty$ given we choose $\LF C_k^{\pm},\, D_k^{\pm} \RF = \LF 1,\,0 \RF $ compatible with the Wronskian
 \begin{equation}
 v_{\pm,k}v_{\pm,k}^{\prime\ast}-v_{\pm,k}^\ast v_{\pm,k}^{\prime}=\pm \:i  \:\:(\implies \vert C_k^{\pm}\vert^2-\vert D_k^{\pm}\vert^2=1). \end{equation}
One can notice that the Wronskian condition equating to $-i$ corresponds to the canonical commutation relation for a reversed arrow of time \cite{Donoghue:2019ecz}
\begin{equation}
	\Big[ \hat{v}_{-}\LF -\tau,\,-\textbf{x} \RF,\,  \hat{\Pi}_{-}\LF -\tau,\,-\textbf{x}^\prime \RF \Big] = -i \delta\LF \textbf{x}-\textbf{x}^\prime \RF\,. 
\end{equation}
Now we can expand \eqref{new-vac1} up to the leading order in $\LF \epsilon, \eta \RF$ as
\begin{equation}
	\begin{aligned}
		{v}_{\pm,\,k} &  \approx \sqrt{\frac{1}{2k}} e^{\pm ik\tau}\LF 1\pm\frac{i}{k\tau} \RF \\ & 
		\pm \LF \epsilon+\frac{\eta}{2} \RF \frac{\sqrt{\pi}}{2\sqrt{k}} \sqrt{\pm k\tau} \frac{\pd H^{(1)}_{\nu_s^{\pm}}\LF \pm k\tau\RF}{\pd \nu_s^{\pm}}\Big\vert_{\nu_s^{\pm}=3/2}
		\label{new-vac1BD}
	\end{aligned}
\end{equation}
We can clearly see from the above equation that $v_{\pm,\,k}$ are not identical but rather differ by the order of slow-roll parameters and scale dependence present in the form of Hankel functions. This exactly implies when a mode $k\sim \frac{1}{r_H}$ exits the horizon, its value would be dictated by $v_{+,\,k}$ at one end $\LF \theta,\, \varphi \RF$ and by $v_{-,\,k}$ on the other end $\LF \pi-\theta,\, \pi+\varphi \RF$. We can derive the analogous result in the vase of inflationary tensor modes well \cite{Kumar:2022zff}. When we compute the equal time correlations of primordial modes on the two opposite points (antipodal) of the three sphere of radius $r_H$ we would see a clear deference between this. This would exactly generate HPA for primordial power spectra.

\section{HPA of primordial power spectra from DQFT of inflatonary quantum fluctuations}

\label{sec:HPA}

\begin{figure}[ht]
	\centering
	\includegraphics[width=0.8\linewidth]{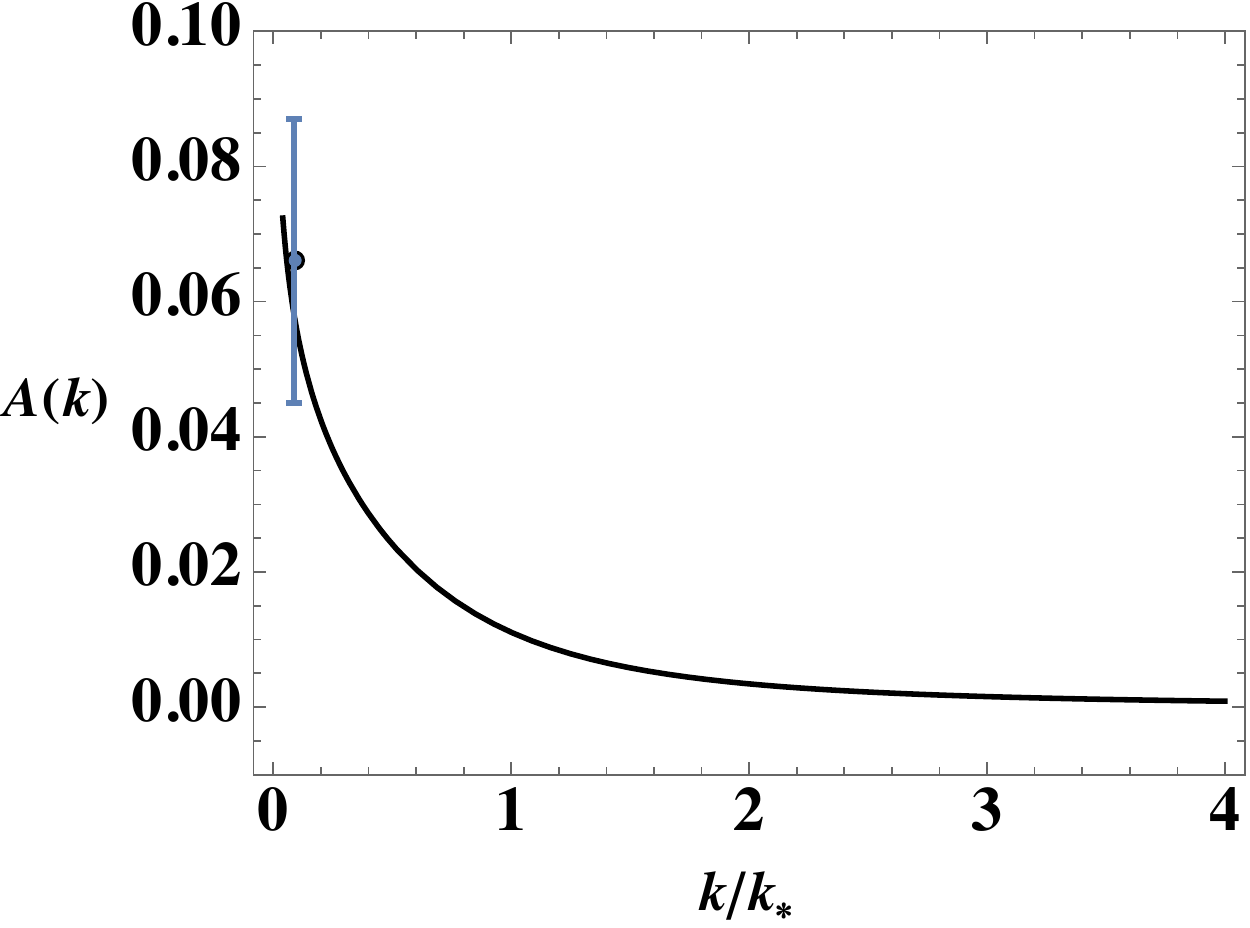}
	\caption{The amplitude of CMB HPA obtained from \eqref{Ako}. In the plot we consider the pivot scale $k_\ast=a_\ast H_\ast=0.05 {\rm Mpc}^{-1}$ and $n_s=0.963$. The blue dot with error bar in the plot corresponds to the observational constraint $\vert A \vert = 0.066\pm 0.021$ on HPA at large angular scales or at low-$\ell-2-64$ or $k\lesssim 10^{-1}k_\ast$.  }
	\label{fig:fig1}
\end{figure}

Inflationary quantum fluctuations in the context of DQFT come with deformations when they exit the comoving horizon on the two opposite sides. 
To calculate the two point correlations of curvature perturbation on super-horizon scales we need to rescale the canonical fields $\hat{v}$ using the classical background quantities as
\begin{equation}
	\begin{aligned}
		{}_{\rm qdS}\langle 0 \vert \zeta_\textbf{k} \zeta_{\textbf{k}^\prime} \vert 0\rangle_{\rm qdS} & =  \LF \frac{1}{2a^2\epsilon}\RF\Bigg\vert_{\rm clas.} 
		\frac{1}{2}\Big[{}_{\rm qdS_I}\langle 0 \vert \hat{v}_{+\,\textbf{k}} \hat{v}_{+\,\textbf{k}^\prime} \vert 0\rangle_{\rm qdS_I}\\ & \quad +{}_{\rm qdS_{II}}\langle 0 \vert \hat{v}_{-\,\textbf{k}} \hat{v}_{-\,\textbf{k}^\prime} \vert 0\rangle_{\rm qdS_{II}}\Big] \\ 
		& = \frac{2\pi^2}{k^3}\LF P_{\zeta_{+}}+P_{\zeta_{-}} \RF \delta\LF \textbf{k}+\textbf{k}^\prime \RF\,, 
	\end{aligned}
	\label{twocor}
\end{equation}
where $\zeta_+,\,\zeta_{-}$ are the power spectra of primordial scalar field in the two hemispheres of the CMB sky. 
{In deriving \eqref{twocor} we must use carefully the direct-sum operation \cite{Conway}}
\begin{equation}
	\hat{v}\vert 0 \rangle_{\rm qdS} = \frac{1}{2}	(\hat{v}_+ \oplus \hat{v}_-) (\vert 0 \rangle_{\rm +} \oplus \vert 0\rangle_{\rm -}) = \frac{1}{2}\begin{pmatrix}
		\hat{v}_+ \vert 0\rangle_{+} \\ 
		\hat{v}_{-} \vert 0 \rangle_{-}\,. 
	\end{pmatrix}
\end{equation} 
The power spectra of curvature perturbation in the two hemispheres are 
\begin{equation}
	\begin{aligned}
		P_{\zeta_{\pm}}&  = \frac{k^3}{2\pi^2}\frac{1}{2a^2\epsilon} \vert v_{\pm,\,k}\vert^2\Bigg\vert_{\tau = \pm \frac{1}{aH}} \\ 
		& \approx \frac{H^2}{8\pi^2\epsilon}  \LF 1+\LF \frac{k}{aH} \RF^2 \RF \\ & 
		\quad \pm \frac{H^2}{4\pi\epsilon} \LF \epsilon+\frac{\eta}{2} \RF \LF\frac{k}{aH}\RF^3 H_{3/2}^{(1)} \:\: \frac{\pd H^{(1)}_{\nu_s^{\pm}}}{\pd \nu_s^{\pm}}\Big\vert_{\nu_s^{\pm}=3/2}\, . 
	\end{aligned}
	\label{pw12}
\end{equation}
%If we combine the above two expressions we obtain 
%\begin{equation}
%	P_\zeta = \frac{1}{2} \LF P_{\zeta_{+}} + P_{\zeta_{-}} \RF  = \frac{H^2}%{8\pi^2\epsilon}\Bigg\vert_{k=aH}
%\end{equation}
%which is the expression in the context of standard inflation.  
Following the meaning of \eqref{Akg} the amplitude of HPA can be deduced as 
\begin{equation}
	A(k) = \frac{P_{\zeta_{+}}-P_{\zeta_{-}}}{4P_\zeta}
	\label{Ako}
\end{equation}
Computing the tilt of the power spectra in the two hemispheres of the CMB sky we get that they are nearly equal. 
\begin{equation}
	\frac{d\ln P_{\zeta_{+}}}{d\ln k}\approx 	\frac{d\ln P_{\zeta_{-}}}{d\ln k} \approx n_s-1 \approx -2\epsilon-\eta\,.
	\label{tilts2}
\end{equation}
This is because of the classical rescaling of the fields in \eqref{twocor} and there would differences in the values of  $n_s$ in the sub-leading order. This can be verified easily by using the expressions of power spectrum \eqref{pw12}. 
Our result is consistent with the observations about $n_s$ in the hemispheres of the CMB. \cite{Mukherjee:2015mma,Axelsson:2013mva}. 
In Fig.~\ref{fig:fig1} we depict the HPA amplitude of the scalar power spectra for $n_s = 0.963$. This is universal result and is valid for any single-field inflation that is compatible with the constraints on $n_s$. 

Similarly, we can apply the DQFT for tensor modes by writing {a tensor fluctuation operator as a direct-sum of two components which describe the tensor mode evolving forward in time at position $\textbf{x}$ and the mode evolving backward in time at position $-\textbf{x}$. 
\begin{equation}
	\hat{u}_{ij} = \frac{1}{\sqrt{2}} \hat{u}^{+}_{ij}\LF \tau,\, \textbf{x} \RF \oplus  \frac{1}{\sqrt{2}}\hat{u}^{-}_{ij}\LF -\tau,\,-\textbf{x} \RF
\end{equation}
Following extremely similar steps it is rather straight forward to compute the tensor mode exiting the comoving horizon on the two opposite sides as
\begin{equation}
	\begin{aligned}
		{u}^{\pm}_{ij,\,k} 
		& \approx e_{ij}\sqrt{\frac{1}{2k}} e^{\pm ik\tau}\LF 1 \pm \frac{i}{k\tau} \RF \\ & 
		\pm e_{ij} \epsilon\frac{\sqrt{\pi}}{2\sqrt{k}} \sqrt{\pm k\tau} \frac{\pd H^{(1)}_{\nu_t^{\pm}}\LF \pm k\tau \RF}{\pd \nu_t^{\pm}}\Big\vert_{\nu_t^{\pm}=3/2}
	\end{aligned}
	\label{new-vac1TBD}
\end{equation}
where $e_{ij}$ denotes the polarization tensor. 
As in the case of the scalar power spectra, we also obtain here two tensor power spectra which describe two point tensor correlations in the direction $\hat{\textbf{n}}$ and $-\hat{\textbf{n}}$ respectively. Thus the tensor power spectra in the two hemispheres of the primordial gravitational wave sky is give by
\begin{equation}
	\begin{aligned}
		P_{h_{\pm}} & = \frac{k^3}{2\pi^2}\frac{4}{a^2} \vert {u}^{\pm}_{ij,\,k}\vert^2\Bigg\vert_{\tau = \pm\frac{1}{aH}} \\ 
		& \approx \frac{2H^2}{\pi^2} \LF 1+\LF \frac{k}{aH} \RF^2 \RF \\ 
		&\quad \pm \frac{H^2}{\pi}\epsilon\LF\frac{k}{aH}\RF^3 H_{3/2}^{(1)} \:\: \frac{\pd H^{(1)}_{\nu_t^{\pm}}}{\pd \nu_t^{\pm}}\Big\vert_{\nu_t^{\pm}=3/2}\, . 
	\end{aligned}
	\label{pwt12}
\end{equation}
Very similar to \eqref{Ako} we can define the HPA of tensor power spectrum as 
\begin{equation}
	T(k) = \frac{P_{h_{+}}-P_{h_{-}}}{4P_h}
	\label{Tko}
\end{equation}
We can quantify the above amplitude $T(k)$ given a model of inflation. 
%in Einstein frame 
%\begin{equation}
%	V(\phi) = \frac{\lambda}{4} \LF 1- e^{-\sqrt{\frac{2}{3}}\phi} \RF^2\,.
%	\label{staro-in}
%\end{equation}
\cite{Kehagias:2013mya}
%&\begin{equation}
%	\epsilon = \frac{3}{4N^2},\quad \eta= \frac{2}{N}
%\end{equation}
In Fig.~\ref{fig:fig2} we depict HPA amplitude of the tensor power spectra for the case of Starobinsky or Higgs inflation. 
\begin{figure}[ht]
	\centering
	\includegraphics[width=0.8\linewidth]{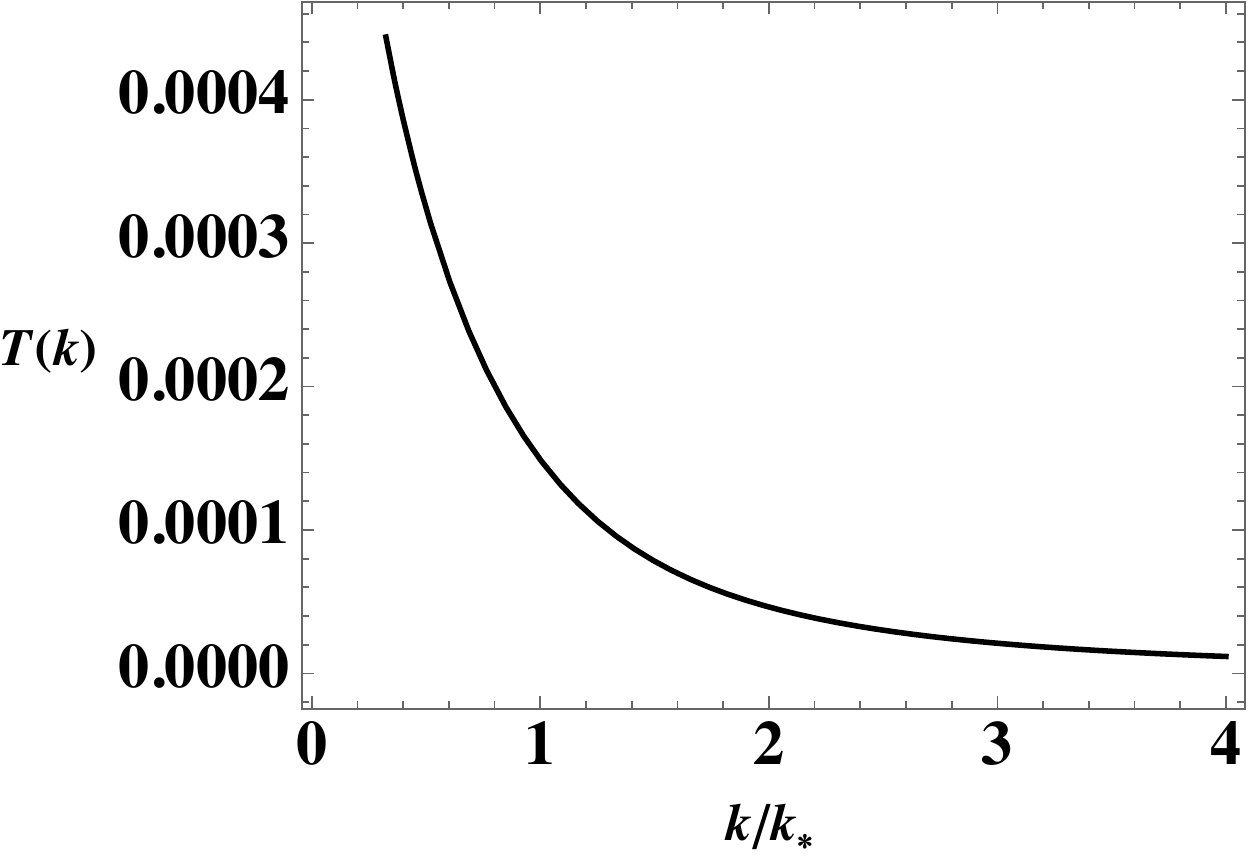}
	\caption{The amplitude of HPA for tensor power spectra obtained from \eqref{Tko}. In the above plot we consider Starobinsky or Higgs inflation for $N=55$ number of e-folds corresponding to the pivot scale $k_\ast =0.05 {\rm Mpc}^{-1}$.}
	\label{fig:fig2}
\end{figure}
Like the scalar power spectra tilt \eqref{tilts2}, we also obtain the same tilt of the tensor power spectra  in the two opposite directions of the sky i.e.,
\begin{equation}
	\frac{d\ln P_{h_{+}}}{d\ln k}\approx 	\frac{d\ln P_{h_{-}}}{d\ln k} \approx n_t \approx -2\epsilon\,.
	\label{tilts3}
\end{equation}
From Fig.~\ref{fig:fig1} and Fig.~\ref{fig:fig2} we can notice that the HPA decreases as the wavenumber increases. This is very much a reasonable result, short wavelength modes would feel less the curvature of spacetime. Both Fig.~\ref{fig:fig1} and Fig.~\ref{fig:fig2} indicate a spontaneous breaking of $\Cc\Pc\Tc$ if confirmed we would transform our understanding of quantum fields in curved spacetime. 

\section{Conclusions}
\label{sec:conc}

In this paper we discussed recent result of \cite{Kumar:2022zff} hemispherical power asymmetry of primordial power spectra which has emerged from fundamental questions of discrete spacetime transformations and their role in quantizing fields in curved spacetime. The result is based on a new scheme of quantizing fields in curved spacetime especially in inflationary spacetime where field operators and Fock space are proposed as direct-sum of two components defined by $\Cc\Pc\Tc$ transformations in the gravitational context. This method of quantization is called direct-sum quantum field theory (DQFT) and it gives us a way to fully exploit the discrete spacetime transformations which form the crux of quantum thoery. 
If the HPA of scalar and tensor power spectra according to DQFT are confirmed in the future observations of CMB and primordial gravitational waves, we would learn more clearly about quantum origin of inflationary fluctuations. It is worth to stress here that HPA of tensor power spectra is predicted for the first time ever based on pure theoretical grounds. 
 It is obviously a natural step to explore further DQFT and also compute higher order correlations functions within this program because as a matter of interest the recent Planck data \cite{Akrami:2019bkn} appears to indicate presence of HPA even in the higher order correlations. 

\begin{acknowledgments}
	KSK acknowledges the support from JSPS and KAKENHI Grant-in-Aid for Scientific Research No. JP20F20320 and No. JP21H00069, and thank Mainz U. for hospitality where part of the work has been carried out. KSK would also like to thank Royal Society for the Newton International Fellowship. 
	 J. Marto is supported by the grant UIDB/MAT/00212/2020.  KSK would like to thank the organizers of \say{Tensions in cosmology 2022} for the kind invitation to present this talk. 
\end{acknowledgments}

\bibliographystyle{apsrev4-2}
\bibliography{corfu.bib}

\begin{thebibliography}{22}%
\makeatletter
\providecommand \@ifxundefined [1]{%
 \@ifx{#1\undefined}
}%
\providecommand \@ifnum [1]{%
 \ifnum #1\expandafter \@firstoftwo
 \else \expandafter \@secondoftwo
 \fi
}%
\providecommand \@ifx [1]{%
 \ifx #1\expandafter \@firstoftwo
 \else \expandafter \@secondoftwo
 \fi
}%
\providecommand \natexlab [1]{#1}%
\providecommand \enquote  [1]{``#1''}%
\providecommand \bibnamefont  [1]{#1}%
\providecommand \bibfnamefont [1]{#1}%
\providecommand \citenamefont [1]{#1}%
\providecommand \href@noop [0]{\@secondoftwo}%
\providecommand \href [0]{\begingroup \@sanitize@url \@href}%
\providecommand \@href[1]{\@@startlink{#1}\@@href}%
\providecommand \@@href[1]{\endgroup#1\@@endlink}%
\providecommand \@sanitize@url [0]{\catcode `\\12\catcode `\$12\catcode
  `\&12\catcode `\#12\catcode `\^12\catcode `\_12\catcode `\%12\relax}%
\providecommand \@@startlink[1]{}%
\providecommand \@@endlink[0]{}%
\providecommand \url  [0]{\begingroup\@sanitize@url \@url }%
\providecommand \@url [1]{\endgroup\@href {#1}{\urlprefix }}%
\providecommand \urlprefix  [0]{URL }%
\providecommand \Eprint [0]{\href }%
\providecommand \doibase [0]{https://doi.org/}%
\providecommand \selectlanguage [0]{\@gobble}%
\providecommand \bibinfo  [0]{\@secondoftwo}%
\providecommand \bibfield  [0]{\@secondoftwo}%
\providecommand \translation [1]{[#1]}%
\providecommand \BibitemOpen [0]{}%
\providecommand \bibitemStop [0]{}%
\providecommand \bibitemNoStop [0]{.\EOS\space}%
\providecommand \EOS [0]{\spacefactor3000\relax}%
\providecommand \BibitemShut  [1]{\csname bibitem#1\endcsname}%
\let\auto@bib@innerbib\@empty
%</preamble>
\bibitem [{\citenamefont {Starobinsky}(1980)}]{Starobinsky:1980te}%
  \BibitemOpen
  \bibfield  {author} {\bibinfo {author} {\bibfnamefont {A.~A.}\ \bibnamefont
  {Starobinsky}},\ }\href {https://doi.org/10.1016/0370-2693(80)90670-X}
  {\bibfield  {journal} {\bibinfo  {journal} {Phys. Lett.}\ }\textbf {\bibinfo
  {volume} {B91}},\ \bibinfo {pages} {99} (\bibinfo {year} {1980})}\BibitemShut
  {NoStop}%
%%CITATION = PHLTA,B91,99;%%
\bibitem [{\citenamefont {Starobinsky}(1979)}]{Starobinsky:1979ty}%
  \BibitemOpen
  \bibfield  {author} {\bibinfo {author} {\bibfnamefont {A.~A.}\ \bibnamefont
  {Starobinsky}},\ }\href@noop {} {\bibfield  {journal} {\bibinfo  {journal}
  {JETP Lett.}\ }\textbf {\bibinfo {volume} {30}},\ \bibinfo {pages} {682}
  (\bibinfo {year} {1979})},\ \bibinfo {note} {[Pisma Zh. Eksp. Teor.
  Fiz.30,719(1979)]}\BibitemShut {NoStop}%
%%CITATION = JTPLA,30,682;%%
\bibitem [{\citenamefont {Akrami}\ \emph {et~al.}(2018)\citenamefont {Akrami}
  \emph {et~al.}}]{Akrami:2018odb}%
  \BibitemOpen
  \bibfield  {author} {\bibinfo {author} {\bibfnamefont {Y.}~\bibnamefont
  {Akrami}} \emph {et~al.} (\bibinfo {collaboration} {Planck}),\ }\href@noop {}
  {\  (\bibinfo {year} {2018})},\ \Eprint {https://arxiv.org/abs/1807.06211}
  {arXiv:1807.06211 [astro-ph.CO]} \BibitemShut {NoStop}%
%%CITATION = ARXIV:1807.06211;%%
\bibitem [{\citenamefont {Akrami}\ \emph {et~al.}(2019)\citenamefont {Akrami}
  \emph {et~al.}}]{Akrami:2019bkn}%
  \BibitemOpen
  \bibfield  {author} {\bibinfo {author} {\bibfnamefont {Y.}~\bibnamefont
  {Akrami}} \emph {et~al.} (\bibinfo {collaboration} {Planck}),\ }\href@noop {}
  {\  (\bibinfo {year} {2019})},\ \Eprint {https://arxiv.org/abs/1906.02552}
  {arXiv:1906.02552 [astro-ph.CO]} \BibitemShut {NoStop}%
\bibitem [{\citenamefont {Yeung}\ and\ \citenamefont
  {Chu}(2022)}]{Yeung:2022smn}%
  \BibitemOpen
  \bibfield  {author} {\bibinfo {author} {\bibfnamefont {S.}~\bibnamefont
  {Yeung}}\ and\ \bibinfo {author} {\bibfnamefont {M.-C.}\ \bibnamefont
  {Chu}},\ }\href {https://doi.org/10.1103/PhysRevD.105.083508} {\bibfield
  {journal} {\bibinfo  {journal} {Phys. Rev. D}\ }\textbf {\bibinfo {volume}
  {105}},\ \bibinfo {pages} {083508} (\bibinfo {year} {2022})},\ \Eprint
  {https://arxiv.org/abs/2201.03799} {arXiv:2201.03799 [astro-ph.CO]}
  \BibitemShut {NoStop}%
\bibitem [{\citenamefont {Aluri}\ \emph {et~al.}(2022)\citenamefont {Aluri}
  \emph {et~al.}}]{Aluri:2022hzs}%
  \BibitemOpen
  \bibfield  {author} {\bibinfo {author} {\bibfnamefont {P.~K.}\ \bibnamefont
  {Aluri}} \emph {et~al.},\ }\href@noop {} {\  (\bibinfo {year} {2022})},\
  \Eprint {https://arxiv.org/abs/2207.05765} {arXiv:2207.05765 [astro-ph.CO]}
  \BibitemShut {NoStop}%
\bibitem [{\citenamefont {Akrami}\ \emph {et~al.}(2014)\citenamefont {Akrami},
  \citenamefont {Fantaye}, \citenamefont {Shafieloo}, \citenamefont {Eriksen},
  \citenamefont {Hansen}, \citenamefont {Banday},\ and\ \citenamefont
  {G\'orski}}]{Akrami:2014eta}%
  \BibitemOpen
  \bibfield  {author} {\bibinfo {author} {\bibfnamefont {Y.}~\bibnamefont
  {Akrami}}, \bibinfo {author} {\bibfnamefont {Y.}~\bibnamefont {Fantaye}},
  \bibinfo {author} {\bibfnamefont {A.}~\bibnamefont {Shafieloo}}, \bibinfo
  {author} {\bibfnamefont {H.~K.}\ \bibnamefont {Eriksen}}, \bibinfo {author}
  {\bibfnamefont {F.~K.}\ \bibnamefont {Hansen}}, \bibinfo {author}
  {\bibfnamefont {A.~J.}\ \bibnamefont {Banday}},\ and\ \bibinfo {author}
  {\bibfnamefont {K.~M.}\ \bibnamefont {G\'orski}},\ }\href
  {https://doi.org/10.1088/2041-8205/784/2/L42} {\bibfield  {journal} {\bibinfo
   {journal} {Astrophys. J. Lett.}\ }\textbf {\bibinfo {volume} {784}},\
  \bibinfo {pages} {L42} (\bibinfo {year} {2014})},\ \Eprint
  {https://arxiv.org/abs/1402.0870} {arXiv:1402.0870 [astro-ph.CO]}
  \BibitemShut {NoStop}%
\bibitem [{\citenamefont {Creswell}\ and\ \citenamefont
  {Naselsky}(2021)}]{Creswell:2021eqi}%
  \BibitemOpen
  \bibfield  {author} {\bibinfo {author} {\bibfnamefont {J.}~\bibnamefont
  {Creswell}}\ and\ \bibinfo {author} {\bibfnamefont {P.}~\bibnamefont
  {Naselsky}},\ }\href {https://doi.org/10.1088/1475-7516/2021/03/103}
  {\bibfield  {journal} {\bibinfo  {journal} {JCAP}\ }\textbf {\bibinfo
  {volume} {03}},\ \bibinfo {pages} {103}},\ \Eprint
  {https://arxiv.org/abs/2102.13442} {arXiv:2102.13442 [astro-ph.CO]}
  \BibitemShut {NoStop}%
\bibitem [{\citenamefont {Schwarz}\ \emph {et~al.}(2016)\citenamefont
  {Schwarz}, \citenamefont {Copi}, \citenamefont {Huterer},\ and\ \citenamefont
  {Starkman}}]{Schwarz:2015cma}%
  \BibitemOpen
  \bibfield  {author} {\bibinfo {author} {\bibfnamefont {D.~J.}\ \bibnamefont
  {Schwarz}}, \bibinfo {author} {\bibfnamefont {C.~J.}\ \bibnamefont {Copi}},
  \bibinfo {author} {\bibfnamefont {D.}~\bibnamefont {Huterer}},\ and\ \bibinfo
  {author} {\bibfnamefont {G.~D.}\ \bibnamefont {Starkman}},\ }\href
  {https://doi.org/10.1088/0264-9381/33/18/184001} {\bibfield  {journal}
  {\bibinfo  {journal} {Class. Quant. Grav.}\ }\textbf {\bibinfo {volume}
  {33}},\ \bibinfo {pages} {184001} (\bibinfo {year} {2016})},\ \Eprint
  {https://arxiv.org/abs/1510.07929} {arXiv:1510.07929 [astro-ph.CO]}
  \BibitemShut {NoStop}%
\bibitem [{\citenamefont {Ade}\ \emph {et~al.}(2014)\citenamefont {Ade} \emph
  {et~al.}}]{Planck:2013lks}%
  \BibitemOpen
  \bibfield  {author} {\bibinfo {author} {\bibfnamefont {P.~A.~R.}\
  \bibnamefont {Ade}} \emph {et~al.} (\bibinfo {collaboration} {Planck}),\
  }\href {https://doi.org/10.1051/0004-6361/201321534} {\bibfield  {journal}
  {\bibinfo  {journal} {Astron. Astrophys.}\ }\textbf {\bibinfo {volume}
  {571}},\ \bibinfo {pages} {A23} (\bibinfo {year} {2014})},\ \Eprint
  {https://arxiv.org/abs/1303.5083} {arXiv:1303.5083 [astro-ph.CO]}
  \BibitemShut {NoStop}%
\bibitem [{\citenamefont {Hansen}\ \emph {et~al.}(2009)\citenamefont {Hansen},
  \citenamefont {Banday}, \citenamefont {Gorski}, \citenamefont {Eriksen},\
  and\ \citenamefont {Lilje}}]{Hansen:2008ym}%
  \BibitemOpen
  \bibfield  {author} {\bibinfo {author} {\bibfnamefont {F.~K.}\ \bibnamefont
  {Hansen}}, \bibinfo {author} {\bibfnamefont {A.~J.}\ \bibnamefont {Banday}},
  \bibinfo {author} {\bibfnamefont {K.~M.}\ \bibnamefont {Gorski}}, \bibinfo
  {author} {\bibfnamefont {H.~K.}\ \bibnamefont {Eriksen}},\ and\ \bibinfo
  {author} {\bibfnamefont {P.~B.}\ \bibnamefont {Lilje}},\ }\href
  {https://doi.org/10.1088/0004-637X/704/2/1448} {\bibfield  {journal}
  {\bibinfo  {journal} {Astrophys. J.}\ }\textbf {\bibinfo {volume} {704}},\
  \bibinfo {pages} {1448} (\bibinfo {year} {2009})},\ \Eprint
  {https://arxiv.org/abs/0812.3795} {arXiv:0812.3795 [astro-ph]} \BibitemShut
  {NoStop}%
\bibitem [{\citenamefont {Aiola}\ \emph {et~al.}(2015)\citenamefont {Aiola},
  \citenamefont {Wang}, \citenamefont {Kosowsky}, \citenamefont
  {Kahniashvili},\ and\ \citenamefont {Firouzjahi}}]{Aiola:2015rqa}%
  \BibitemOpen
  \bibfield  {author} {\bibinfo {author} {\bibfnamefont {S.}~\bibnamefont
  {Aiola}}, \bibinfo {author} {\bibfnamefont {B.}~\bibnamefont {Wang}},
  \bibinfo {author} {\bibfnamefont {A.}~\bibnamefont {Kosowsky}}, \bibinfo
  {author} {\bibfnamefont {T.}~\bibnamefont {Kahniashvili}},\ and\ \bibinfo
  {author} {\bibfnamefont {H.}~\bibnamefont {Firouzjahi}},\ }\href
  {https://doi.org/10.1103/PhysRevD.92.063008} {\bibfield  {journal} {\bibinfo
  {journal} {Phys. Rev. D}\ }\textbf {\bibinfo {volume} {92}},\ \bibinfo
  {pages} {063008} (\bibinfo {year} {2015})},\ \Eprint
  {https://arxiv.org/abs/1506.04405} {arXiv:1506.04405 [astro-ph.CO]}
  \BibitemShut {NoStop}%
\bibitem [{\citenamefont {Lyth}(2013)}]{Lyth:2013vha}%
  \BibitemOpen
  \bibfield  {author} {\bibinfo {author} {\bibfnamefont {D.~H.}\ \bibnamefont
  {Lyth}},\ }\href {https://doi.org/10.1088/1475-7516/2013/08/007} {\bibfield
  {journal} {\bibinfo  {journal} {JCAP}\ }\textbf {\bibinfo {volume} {08}},\
  \bibinfo {pages} {007}},\ \Eprint {https://arxiv.org/abs/1304.1270}
  {arXiv:1304.1270 [astro-ph.CO]} \BibitemShut {NoStop}%
\bibitem [{\citenamefont {Lyth}(2015)}]{Lyth:2014mga}%
  \BibitemOpen
  \bibfield  {author} {\bibinfo {author} {\bibfnamefont {D.~H.}\ \bibnamefont
  {Lyth}},\ }\href {https://doi.org/10.1088/1475-7516/2015/04/039} {\bibfield
  {journal} {\bibinfo  {journal} {JCAP}\ }\textbf {\bibinfo {volume} {04}},\
  \bibinfo {pages} {039}},\ \Eprint {https://arxiv.org/abs/1405.3562}
  {arXiv:1405.3562 [astro-ph.CO]} \BibitemShut {NoStop}%
\bibitem [{\citenamefont {Sravan~Kumar}\ and\ \citenamefont {Marto}()}]{KSKJM}%
  \BibitemOpen
  \bibfield  {author} {\bibinfo {author} {\bibfnamefont {K.}~\bibnamefont
  {Sravan~Kumar}}\ and\ \bibinfo {author} {\bibfnamefont {J.}~\bibnamefont
  {Marto}},\ }\href@noop {} {\ }\Eprint {https://arxiv.org/abs/2305.XXXX}
  {arXiv:2305.XXXX [hep-th]} \BibitemShut {NoStop}%
\bibitem [{\citenamefont {Kumar}\ and\ \citenamefont
  {Marto}(2022)}]{Kumar:2022zff}%
  \BibitemOpen
  \bibfield  {author} {\bibinfo {author} {\bibfnamefont {K.~S.}\ \bibnamefont
  {Kumar}}\ and\ \bibinfo {author} {\bibfnamefont {J.~a.}\ \bibnamefont
  {Marto}},\ }\href@noop {} {\  (\bibinfo {year} {2022})},\ \Eprint
  {https://arxiv.org/abs/2209.03928} {arXiv:2209.03928 [gr-qc]} \BibitemShut
  {NoStop}%
\bibitem [{\citenamefont {Conway}(2010)}]{Conway}%
  \BibitemOpen
  \bibfield  {author} {\bibinfo {author} {\bibfnamefont {J.~B.}\ \bibnamefont
  {Conway}},\ }\href@noop {} {\emph {\bibinfo {title} {A course in functional
  analysis}}},\ \bibinfo {edition} {2nd}\ ed.,\ Graduate texts in mathematics ;
  96\ (\bibinfo  {publisher} {Springer Science+Business Media},\ \bibinfo
  {address} {New York},\ \bibinfo {year} {2010})\BibitemShut {NoStop}%
\bibitem [{\citenamefont {Rovelli}(2004)}]{Rovelli:2004tv}%
  \BibitemOpen
  \bibfield  {author} {\bibinfo {author} {\bibfnamefont {C.}~\bibnamefont
  {Rovelli}},\ }\href {https://doi.org/10.1017/CBO9780511755804} {\emph
  {\bibinfo {title} {{Quantum gravity}}}},\ Cambridge Monographs on
  Mathematical Physics\ (\bibinfo  {publisher} {Univ. Pr.},\ \bibinfo {address}
  {Cambridge, UK},\ \bibinfo {year} {2004})\BibitemShut {NoStop}%
\bibitem [{\citenamefont {Donoghue}\ and\ \citenamefont
  {Menezes}(2019)}]{Donoghue:2019ecz}%
  \BibitemOpen
  \bibfield  {author} {\bibinfo {author} {\bibfnamefont {J.~F.}\ \bibnamefont
  {Donoghue}}\ and\ \bibinfo {author} {\bibfnamefont {G.}~\bibnamefont
  {Menezes}},\ }\href {https://doi.org/10.1103/PhysRevLett.123.171601}
  {\bibfield  {journal} {\bibinfo  {journal} {Phys. Rev. Lett.}\ }\textbf
  {\bibinfo {volume} {123}},\ \bibinfo {pages} {171601} (\bibinfo {year}
  {2019})},\ \Eprint {https://arxiv.org/abs/1908.04170} {arXiv:1908.04170
  [hep-th]} \BibitemShut {NoStop}%
\bibitem [{\citenamefont {Mukherjee}\ \emph {et~al.}(2016)\citenamefont
  {Mukherjee}, \citenamefont {Aluri}, \citenamefont {Das}, \citenamefont
  {Shaikh},\ and\ \citenamefont {Souradeep}}]{Mukherjee:2015mma}%
  \BibitemOpen
  \bibfield  {author} {\bibinfo {author} {\bibfnamefont {S.}~\bibnamefont
  {Mukherjee}}, \bibinfo {author} {\bibfnamefont {P.~K.}\ \bibnamefont
  {Aluri}}, \bibinfo {author} {\bibfnamefont {S.}~\bibnamefont {Das}}, \bibinfo
  {author} {\bibfnamefont {S.}~\bibnamefont {Shaikh}},\ and\ \bibinfo {author}
  {\bibfnamefont {T.}~\bibnamefont {Souradeep}},\ }\href
  {https://doi.org/10.1088/1475-7516/2016/06/042} {\bibfield  {journal}
  {\bibinfo  {journal} {JCAP}\ }\textbf {\bibinfo {volume} {06}},\ \bibinfo
  {pages} {042}},\ \Eprint {https://arxiv.org/abs/1510.00154} {arXiv:1510.00154
  [astro-ph.CO]} \BibitemShut {NoStop}%
\bibitem [{\citenamefont {Axelsson}\ \emph {et~al.}(2013)\citenamefont
  {Axelsson}, \citenamefont {Fantaye}, \citenamefont {Hansen}, \citenamefont
  {Banday}, \citenamefont {Eriksen},\ and\ \citenamefont
  {Gorski}}]{Axelsson:2013mva}%
  \BibitemOpen
  \bibfield  {author} {\bibinfo {author} {\bibfnamefont {M.}~\bibnamefont
  {Axelsson}}, \bibinfo {author} {\bibfnamefont {Y.}~\bibnamefont {Fantaye}},
  \bibinfo {author} {\bibfnamefont {F.~K.}\ \bibnamefont {Hansen}}, \bibinfo
  {author} {\bibfnamefont {A.~J.}\ \bibnamefont {Banday}}, \bibinfo {author}
  {\bibfnamefont {H.~K.}\ \bibnamefont {Eriksen}},\ and\ \bibinfo {author}
  {\bibfnamefont {K.~M.}\ \bibnamefont {Gorski}},\ }\href
  {https://doi.org/10.1088/2041-8205/773/1/L3} {\bibfield  {journal} {\bibinfo
  {journal} {Astrophys. J. Lett.}\ }\textbf {\bibinfo {volume} {773}},\
  \bibinfo {pages} {L3} (\bibinfo {year} {2013})},\ \Eprint
  {https://arxiv.org/abs/1303.5371} {arXiv:1303.5371 [astro-ph.CO]}
  \BibitemShut {NoStop}%
\bibitem [{\citenamefont {Kehagias}\ \emph {et~al.}(2014)\citenamefont
  {Kehagias}, \citenamefont {Moradinezhad~Dizgah},\ and\ \citenamefont
  {Riotto}}]{Kehagias:2013mya}%
  \BibitemOpen
  \bibfield  {author} {\bibinfo {author} {\bibfnamefont {A.}~\bibnamefont
  {Kehagias}}, \bibinfo {author} {\bibfnamefont {A.}~\bibnamefont
  {Moradinezhad~Dizgah}},\ and\ \bibinfo {author} {\bibfnamefont
  {A.}~\bibnamefont {Riotto}},\ }\href
  {https://doi.org/10.1103/PhysRevD.89.043527} {\bibfield  {journal} {\bibinfo
  {journal} {Phys. Rev.}\ }\textbf {\bibinfo {volume} {D89}},\ \bibinfo {pages}
  {043527} (\bibinfo {year} {2014})},\ \Eprint
  {https://arxiv.org/abs/1312.1155} {arXiv:1312.1155 [hep-th]} \BibitemShut
  {NoStop}%
%%CITATION = ARXIV:1312.1155;%%
\end{thebibliography}%

\end{document}